\documentclass[12pt]{article}

 \newread\testifexists
 \def\GetIfExists #1 {\immediate\openin\testifexists=#1
     \ifeof\testifexists\immediate\closein\testifexists\else
     \immediate\closein\testifexists\input #1\fi}

 \usepackage{gthstyle}\usepackage{amsfonts}
 \usepackage{amssymb}
 \immediate\openin\testifexists=e
     \ifeof\testifexists\immediate\closein\testifexists\else
     \immediate\closein\testifexists\input e\fipsf

 \def\Bbb#1{\setbox0=\hbox{$\tt #1$}  \copy0\kern-\wd0\kern .1em\copy0}

 \def\bbf#1{\setbox0=\hbox{$#1$} \kern-.025em\copy0\kern-\wd0
         \kern.05em\copy0\kern-\wd0 \kern-.025em\raise.0433em\box0}

 \immediate\openin\testifexists=a
     \ifeof\testifexists\immediate\closein\testifexists\else
     \immediate\closein\testifexists\input a\fimssym.def  

 \def\a{\alpha}      \def\b{\beta}   \def\g{\gamma}      \def\G{\Gamma}
 \def\d{\delta}      \def\D{\Delta}  \def\e{\varepsilon} 
 \def\k{\kappa}      \def\l{\lambda} \def\L{\Lambda}     \def\m{\mu}
 \def\f{\phi}            \def\vv{\varphi}    \def\n{\nu}
 \def\j{\psi}                 \def\s{\sigma}

 \def\w{\omega}        

    \def\LL{{\mathcal L}} \def\OO{{\mathcal O}} \def\DD{{\mathcal D}}
 \def\pa{\partial} \def\ra{\rightarrow}
  
 \def\dd{{\rm d}}  \def\bra{\langle}   \def\ket{\rangle}

 \def\qu{\ {\buildrel {\displaystyle ?} \over =}\ }
 \def\deff{\ {\buildrel{\rm def}\over{=}}\ }
 \def\iss{\ =\ }

 \def\fract#1#2{{\textstyle{#1\over#2}}}
 \def\ffract#1#2{\raise .2 em\hbox{$\scriptstyle#1$}\kern-.3em/
                 \kern-.2em\lower .15 em \hbox{$\scriptstyle#2$}}
 
 \def\half{\fract12} \def\quart{\fract14} 

 \def\ex#1{e^{\textstyle#1}}

               \newcommand{\Tr}{{\mbox{Tr}}\,}
                     \newcommand{\fn}{\footnote}
 \newcommand{\nn}{\nonumber\\[2pt]}             \newcommand{\nm}{\nonumber}
 \newcommand{\be}{\begin{eqnarray}}             \newcommand{\ee}{\end{eqnarray}}
 \newcommand{\bi}[1]{\begin{itemize}\item[#1]}         \newcommand{\itm}[1]{\item[#1]}
       \newcommand{\ei}{\end{itemize}}
 \newcommand{\eqn}[1]{(\ref{#1})}

 \newcommand{\crlb}[1]{\label{#1}\\[2pt]}
 \newcommand{\eela}[1]{\quad\hbox{\scriptsize{#1}}\label{#1}\end{eqnarray}}
 \newcommand{\eelb}[1]{\label{#1}\end{eqnarray}}
 
 \newcommand{\newsecb}[2]{\section{#1}\label{#2}\setcounter{equation}{0}}
 
 \newcommand{\nolabels} {\def\eel{\eelb} \def\crl{\crlb} \def\newsecl{\newsecb}}

\newcommand\publishversion{\nolabels\setlength{\textheight}{9in}\setlength{\oddsidemargin}{0in}
    \setlength{\textwidth}{6.3in}\setlength{\topmargin}{-0.1in}}

\publishversion
\begin{document} \begin{titlepage}

\title{\normalsize \hfill ITP-UU-10/30 \\ \hfill SPIN-10/25
\vskip 20mm \Large\bf Probing the small distance structure of canonical quantum gravity using the conformal
group}

\author{Gerard 't~Hooft}
\date{\normalsize Institute for Theoretical Physics \\
Utrecht University \\ and
\medskip \\ Spinoza Institute \\ Postbox 80.195 \\ 3508 TD Utrecht, the Netherlands \smallskip \\
e-mail: \tt g.thooft@uu.nl \\ internet: \tt http://www.phys.uu.nl/\~{}thooft/}

\maketitle

\begin{quotation} \noindent {\large\bf Abstract } \medskip

{\small{In canonical quantum gravity, the formal functional integral includes an integration over the local
conformal factor, and we propose to perform the functional integral over this factor before doing any of the other
functional integrals. By construction, the resulting effective theory would be expected to be conformally
invariant and therefore finite. However, also the conformal integral itself diverges, and the effects of a
renormalization counter term are considered. It generates problems such as unitarity violation, due to a
Landau-like ghost, and conformal anomalies. Adding (massive or massless) matter fields does not change the
picture. Various alternative ideas are offered, including a more daring speculation, which is that no counter term
should be allowed for at all. This has far-reaching and important consequences, which we discuss. A surprising
picture emerges of quantized elementary particles interacting with a gravitational field, in particular gravitons,
which are ``partly classical''. This approach was inspired by a search towards the reconciliation of Hawking
radiation with unitarity and locality, and it offers basic new insights there.}}

\end{quotation}


\end{titlepage}

\eject
\newsecl{Introduction: splitting the functional integral}{intro}  \def\mat{\mathrm{\,mat}}
The Einstein-Hilbert action of the generally covariant theory of gravity reads
 \be S^{\,\mathrm{total}}=\int\dd^4 x\sqrt{-g}\,\left({\textstyle 1\over 16\pi G_N} R+\LL^\mat\right)\ , \eel{EH}
where the matter Lagrangian \(\mathcal{L}^\mat\) is written in a generally covariant manner using
the space-time metric \(g_{\m\n}(x)\), and \(G_N\) is Newton's constant. In this paper, we begin
studying the case where \(\LL^\mat\) is conformally symmetric, which means that under a space-time
transformation
 \be {x^\m}'=C\,{x^\m-a^\m\over  (x -a)^2}+b^\m ,\ \eel{conftrf}
we have a transformation law for the matter fields such that
 \be && g_{\m\n}(x')=\l(x)^2 g_{\m\n}(x)\ ;\qquad \sqrt{-g(x')}\,{\LL^\mat}'(x')=\sqrt{-g(x)}\,\LL^\mat(x)\
 ; \crl{confS} && {S^\mat}'= S^\mat \
 ,\nn &&  \mathrm{so\ that\ in\ }\,n\,\ \mathrm{dimensions},\qquad {\LL^\mat}'(x')=\l^n\,\LL^\mat(x)\ . \eel{confL}
For the conformal transformation \eqn{conftrf} we have \(\l(x)=C/(x^\m-a^\m)^2\), which leaves flat
spacetime flat, but for curved background space-times, where we drop the condition of flatness,
\(\l(x)\) may be any function of \(x^\m\). There are several examples of such conformally invariant
matter systems such as \(\mathcal{N}=4\) super-Yang-Mills theory in \(n=4\) space-time dimensions.
We will concentrate on \(n=4\).

We begin by temporarily assuming conformal invariance of the matter fields, only for convenience;
later we will see that allowing matter fields to be more general will only slightly modify the
picture.

In canonical gravity, the quantum amplitudes are obtained by functionally integrating the exponent of the entire action
over all components of the metric tensor at all space-time points \(x^\m\):
 \be \G=\int\DD g_{\m\n}(x)\,\DD\vv^\mat(x)\ \ex{iS^\mathrm{\,total}}\ . \eel{functampl}
Although one usually imposes a gauge constraint so as to reduce the size of function space, this is
not necessary formally. In particular, one has to integrate over the common factor \(\w(x)\) of the
metric tensor \(g_{\m\n}(x)\), when we write
 \be g_{\m\n}(x)\deff\w^2(x)\,\hat g_{\m\n}(x)\ , \eel{gsplit}
where \(\hat g_{\m\n}(x)\) may be subject to some arbitrary constraint concerning its overall factor. For instance, in
any coordinate frame one may impose
 \be \det(\hat g)=-1\ , \eel{scalegauge}
\textit{besides} imposing a gauge condition for each of the \(n=4\) coordinates. The quantity
\(\hat g_{\m\n}(x)\) in Eq.~\eqn{scalegauge} does not transform as an ordinary tensor but as what
could be called a ``pseudo"tensor, meaning that it scales unconventionally under coordinate
transformations with a non-trivial Jacobian. \(\w(x)\) is then a ``pseudo"scalar.

The prefix ``pseudo" was put in quotation marks here because, usually, `pseudo' means that the
object receives an extra minus sign under a parity transformation; it is therefore preferred to use
another phrase. For this reason, we replace ``pseudo" by `meta', using the words `metatensor' and
`metascalar' to indicate fields that transform as tensors or scalars, but with prefactors
containing unconventional powers of the Jacobian of the coordinate transformation.

Rewriting
 \be\int\DD g_{\m\n}(x)=\int\DD\w(x)\int\DD\hat g_{\m\n}(x)\ , \eel{functintsplit}
while imposing a gauge constraint\fn{A fine choice would be, for instance, \(\pa_\m\hat g^{\m\n}=0\). Of
course, the usual Faddeev Popov quantization procedure is assumed.} that \emph{only} depends on \(\hat
g_{\m\n}\), not on \(\w\), we now propose \textit{first} to integrate over \(\w(x)\) and then over \(\hat
g_{\m\n}(x)\) and \(\vv^\mat(x)\). This has peculiar consequences, as we will see.

In the standard perturbation expansion, the integration order does not matter. Also, if dimensional
regularization is employed, the choice of the functional metric in the space of all fields \(\w(x)\) and
\(\hat g_{\m\n}(x)\) is unambiguous, as its effects are canceled against all other quartic divergences in the
amplitudes (any ambiguity is represented by integrals of the form \(\int\dd^n k\,\mathrm{Pol}(k)\) which
vanish when dimensionally renormalized).
  \(\w(x)\) acts as a Lagrange multiplier. Again, perturbation expansion tells us how to handle this integral: in
general, \(\w(x)\) has to be chosen to lie on a complex contour. The momentum integrations may be carried out
in Euclidean (Wick rotated) space-time, but even then, \(\w(x)\) must be integrated along a complex contour,
which will later (see Section \ref{scalefint}) be determined to be
 \be\w(x)=1+i\a(x)\ ,\qquad\a\ \hbox{real}. \eel{omegacontour}  If \(\w(x)\) itself
had been chosen real then the Wick rotated functional integral would diverge exponentially so that \(\w\)
would no longer function properly as a Lagrange multiplier.

If there had been no further divergences, one would have expected the following scenario:
 \bi{-} The functional integrand \(\w(x)\) only occurs in the gravitational part of the action, since the matter field
is conformally invariant (non-conformal matter does contribute to this integral, but these would be sub
dominating corrections, see later).
 \itm{-} After integrating over all scale functions \(\w(x)\), but not yet over \(\hat g_{\m\n}\), the resulting effective
action in terms of \(\hat g_{\m\n}\) should be expected to become scale-invariant, \textit{i.e.} if we would split
\(\hat g_{\m\n}\) again as in Eq.~\eqn{gsplit},
 \be \hat g_{\m\n}(x)\qu \hat\w^2(x) \hat{\hat g}_{\m\n}\ , \eel{gdubbelsplit}
no further dependence on \(\hat\w(x)\) should be expected.
 \itm{-} Therefore, the effective action should now describe a conformally invariant theory, both for gravity
and for matter. Because of this, the effective theory might be expected to be renormalizable, or even finite!
If any infinities do remain, one might again employ dimensional renormalization to remove them. \ei
 \noindent However, this expectation is jeopardized by an apparent difficulty: the \(\w\) integration
is indeed ultraviolet divergent.\cite{capperduff} In contrast with the usual procedures in
perturbation theories, it is not associated with an infinitesimal multiplicative constant (such as
the coupling constant in ordinary perturbation theories), and so a renormalization counter term
would actually represent an infinite distortion of the canonical theory. Clearly, renormalization
must be carried out with much more care. Later, in Section~\ref{confaction}, we suggest various
scenarios.

First, the main calculation will be carried out, in the next section. Then, the contributions from conformal
matter fields are considered, and subsequently the effect of non conformal matter, by adding mass terms.
Finally, we will be in a position to ask questions about renormalization (dimensional or otherwise). We end
with conclusions, and an appendix displaying the details of the matter field calculations.

\newsecl{Calculating the divergent part of the scalar functional integral}{scalefint}

Calculations related to the conformal term in gravity, and their associated anomalies, date back from the
early 1970s and have been reviewed amnong others in a nice paper by Duff\cite{Duff}. In particular, we here
focus on footnote (4) in that paper.

First, we go to \(n\) space-time dimensions, in order later to be able to perform dimensional
renormalization. For future convenience (see Eq.~\eqn{EHsplit}), we choose to replace the parameter \(\w\)
then by \(\w^{2/(n-2)}\), so that Eq.~\eqn{gsplit} becomes
 \be g_{\m\n}(x)=\w^\fract 4{n-2}\,\hat g_{\m\n}(x)\ . \eel{gsplitndim}
In terms of \(\hat g_{\m\n}\) and \(\w\), the Einstein-Hilbert action \eqn{EH} now reads
 \be S=\int\dd^nx\sqrt{-\hat g}\left({1\over 16\pi G_N}\bigg(\w^2\hat R+{4(n-1)\over n-2}\,\hat g^{\,\m\n}
 \pa_\m\w\,\pa_\n\w\bigg) +\LL^\mat(\hat g_{\m\n}) \right)\ . \eel{EHsplit}
This shows that the functional integral over the field \(\w(x)\) is a Gaussian one, which can be performed
rigorously: it is a determinant. The conformally invariant matter Lagrangian is independent of \(\w\), but at
a later stage of the theory we shall consider mass terms for matter, which still would allow us to do the
functional integral over \(\w\), but as for now we wish to avoid the associated complications, assuming that,
perhaps, at scales close to the Planck scale the \(\w\) dependence of matter might dwindle.

We use the caret (\(\hat{}\)) to indicate all expressions defined by the metatensor \(\hat
g_{\m\n}\), such as covariant derivatives, as if it were a true tensor.

Note that, in `Euclidean gravity', the \(\w\) integrand has the wrong sign. This is why \(\w\) must be chosen
to be on the contour~\eqn{omegacontour}. In practice, it is easiest to do the functional \(\w\) integration
perturbatively, by writing
 \be \hat g_{\m\n}(x)=\eta_{\m\n}+\k\,h_{\m\n}(x)\ ,\qquad\eta_{\m\n}=\hbox{diag}(-1,1,1,1)\
 ,\qquad\k=\sqrt{8\pi G_N}\ , \eel{etaplush}
and expanding in powers of \(\k\) (although later we will see that that expansion can sometimes be summed). A
factor \(\sqrt{-\hat g}\,8(n-1)/16\pi G_N(n-2)\) in Eq.~\eqn{EHsplit} can be absorbed in the definition of
\(\w\).\fn{Note that, therefore, Newton's constant disappears completely (its use in Eq.~\eqn{etaplush} is
inessential). This a characteristic feature of this approach.} This turns the action \eqn{EHsplit} into
 \be S=\int\dd^nx\sqrt{-\hat g}\left(\half \hat g^{\m\n}\pa_\m\w\pa_\n\w+\half{n-2\over
 4(n-1)}\hat R\w^2+\LL^\mat(\hat g_{\m\n})\right)\ . \eel{omegaL}

Regardless the \(\w\) contour, the \(\w\) propagator can be read off from the action \eqn{omegaL}:
 \be P^{(\w)}(k)=-\,{1\over k^2-i\e }\ , \eel{omegaprop}
where \(k_\m\) is the momentum. The \(i\e\) prescription is the one that follows from the conventional
perturbative theory. We see that there is a kinetic term (perturbed by a possible non-trivial space-time
dependence of \(\hat g_{\m\n}\)), and a direct interaction, ``mass" term proportional to the background
scalar curvature \(\hat R\):
 \be {n-2\over 4(n-1)}\,\hat R\quad \buildrel{n\ra 4}\over{\longrightarrow}\quad \fract16 \hat R\ , \eel{Rterm}

\begin{figure}[h] \setcounter{figure}{0}
\hspace{100pt} \epsfxsize=90 mm\epsfbox{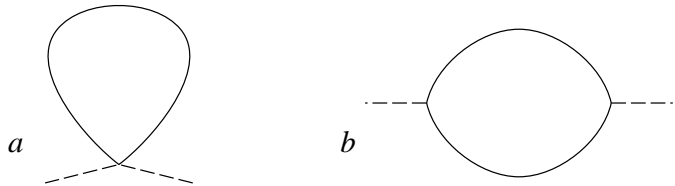}
\begin{center}  \caption{
  Feynman diagrams for the \(\w\) determinant}\label{diag.fig}
\end{center}
\end{figure}

The most important diagrams contributing to the effective action for the remaining field \(\hat
g_{\m\n}\) are the ones indicated in Fig.~\ref{diag.fig}, which include the terms up to
 \(\OO(\k^2)\).  The ``tadpole", Fig.~\ref{diag.fig}a, does not contribute if we apply dimensional
regularization, since there is no mass term in the single propagator that we have,
Eq.~\eqn{omegaprop}. So, in this approximation, we have to deal with the 2-point diagram only.

We can compute the integral
 \be F(q)\deff\int_{\mathrm{Eucl}}{\dd^{n}k\over k^2\,(k-q)^2}
 &=&{\pi^{\half n+\fract 3 2}\,2^{3-n}\,(q^2)^{\half n-2}\over
    \G(\half n-{1\over 2})\,\sin\pi(2-\half n)}\ . \eel{confint1}
Now we will also need integrals containing extra factors \(k_\m\) in the
numerator. Therefore, we define
 \be \bra k\cdots k\ket\deff
{1\over F(q)}\int_{\mathrm{Eucl}}{\dd^{n}k\ k\cdots k\over k^2(k-q)^2}\ .
\eel{kintdef} Then
 \be \bra k_\m\ket&=& \half q_\m\ ;\crl{klinear} \bra k_\m k_\n\ket&=&\fract 1
 {4(n-1)}\bigg(n q_\m q_\n-q^2\d_{\m\n}\bigg)\ ;\\
 \bra k_\m k_\n k_\l\ket &=&\fract 1{8(n-1)}\bigg((n+2)q_\m q_\n
 q_\l-q^2(\d_{\m\n}q_\l+\d_{\n\l}q_\m+\d_{\l\m}q_\n)\bigg)\ ;\\
 \bra k_\m k_\n k_\a k_\b\ket &=&\fract
 1{16(n-1)(n+1)}    \bigg((n+2)(n+4)\,q_\m q_\n q_\a q_\b\nn
 &&- q^2( n+2)(\,\d_{\m\n}q_\a q_\b+[\mathrm{5\ terms}]\,)\nn
 &&+ q^4(\d_{\m\n}\d_{\a\b}+\d_{\m\a}\d_{\n\b}+\d_{\m\b}\d_{\n\a})
  \bigg)  \ , \eel{kintegrals}
where the 5 terms are simply the remaining 5 permutations of the previous
term.

These expressions can now be used to compute all diagrams that contribute to the \(\w\)
determinant, but the calculations are lengthy and not very illuminating. More important are those
parts that diverge as \(n\ra 4\). The expression \eqn{confint1} for \(F(q)\) diverges at \(n\ra
4\), so that all integrals in Eq.~\eqn{kintdef} diverge similarly. By using general covariance, one
can deduce right away that the divergent terms must all combine in such a way that they only depend
on the Riemann curvature. Dimensional arguments then suffice to conclude that the coefficients must
be local expressions in the squares of the curvature.

The key calculations for the divergent parts have already been performed in 1973 \cite{GtHVeltman}.
There, it was found that a Lagrangian of the form
 \be \LL=\sqrt{-g}\left(-\half g^{\m\n}(x)\,\pa_\m\vv\pa_\n\vv+\half M(x)\,\vv^2\right)\ , \eel{tHV}
will generate an effective action, whose divergent part is of the form \def\div{{\mathrm{\,div}}}
 \be{ }\hspace{-20pt} S^\div=\int\dd^nx\,\G^\div(x)\ , \quad \G^\div ={\sqrt{-g}\over
8\pi^2(4-n)}\left(\fract1{120}(R_{\m\n}R^{\m\n}-\fract13 R^2)+ \fract14
 (M+\fract16 R)^2\right)   \eel{scalarpole}
(we use here a slightly modified notation, implying, among others, a sign switch in the definition of the
Ricci curvature, and a minus sign as ref.~\cite{GtHVeltman} calculated the Lagrangian \(\LL+\D \LL,\
 \D \LL=-\G^\div\) needed to obtain a finite theory.)

In our case, we see that, in the Lagrangian \eqn{omegaL} (with the dynamical part of \(\w\)
imaginary),
 \be M=-\fract16\hat R\ ,\qquad\G^\div={\sqrt{-\hat g}\over 960\pi^2(4-n)}(\hat R_{\m\n}\hat R^{\m\n}-\fract13 \hat R^2)\ ,
 \eel{divpart}
since the second term in \eqn{scalarpole} cancels out exactly. Indeed, it had to cancel out, as we will see
shortly.

To see what the divergence here means, we use the fact that the mass dependence of a divergent
integral typically takes the form
 \be C(n)m^{n-4}\G(2-\half n)\ra {C\over 4-n}\left(1+(n-4)\log\big({m\over\L}\big)\right)\ra
 C\bigg(\log\L+{1\over 4-n}\bigg)+\hbox{ finite}\ ,\nn \eel{cutoff}
where \(m\) stands for a mass or an external momentum \(k\), and \(\L\) is some reference mass, such as an ultraviolet
cutoff. Thus, the divergent expression \(1/(4-n)\) generally plays the same role as the logarithm of an ultraviolet
cutoff \(\L\).

\newsecl{Local scale invariance and the Weyl curvature}{Weyl}

\def\eff{\mathrm{eff}}
Assume for a moment that, after having dealt with the divergent expression \eqn{divpart}, the functional
integral over the conformal variable \(\w(x)\) could somehow be made to produce a finite and meaningful
result. We would have a finite effective action \(\G^\eff\) that is completely conformally invariant, and we
would expect to be left with \(\hat g_{\m\n}(x)\) as our remaining dynamical variables.

It is this theory that could be used to handle the black hole complementarity issue. It was explained in
Ref.\cite{GtHcompl} that black hole complementarity means that an observer on his way into a black hole may
experience the surrounding space-time differently from what an outside observer sees. They disagree about the
back reaction from Hawking radiation, and it was argued that this disagreement must include the metascalar
field \(\w(x)\). A completely conformally invariant theory as a starting point could explain this situation;
we return to this issue in Section~\ref{concl}.

Thus, we now consider an effective theory with not only general covariance,
 \be \hat g_{\m\n}\ra \hat g_{\m\n}+\hat D_\m u_\n+\hat D_\n u_\m\ , \eel{infcoordtrf}
where \(u_\m(x)\) are the generators of infinitesimal coordinate transformations, and \(\hat D_\m\)
is the covariant derivative with respect to \(\hat g_{\m\n}\); but now we also have a new kind of
gauge invariance, being local scale invariance, which we write in infinitesimal notation, for
convenience:
 \be \hat g_{\m\n}\ra \hat g_{\m\n}+\l(x) \hat g_{\m\n}\ , \eel{localscale}
and we demand invariance under that as well. Note that this transformation is quite distinct from scale
transformations in the coordinate frame, which of course belongs to \eqn{infcoordtrf} and as such is always
an invariance of the usual theory. In short, we now have a theory with a 5 dimensional local gauge group.
Theories of this sort have been studied in detail\cite{PM}.

The Riemann tensor \(\hat R^\a_{\ \b\m\n}\) transforms as a decent tensor under the coordinate
transformations \eqn{infcoordtrf}, but it is not invariant (or even covariant) under the local scale
transformation \eqn{localscale}. Now, in four space time dimensions, we can split up the 20 independent
components of the Riemann tensor into the 10 component Ricci tensor
 \be \hat R_{\m\n}=\hat R^\a_{\ \m\a\n}\ , \eel{Ricci}
and the components orthogonal to that, called the Weyl tensor,
 \be   &&\hat W_{\m\n\a\b}=\hat R_{\m\n\a\b}+\nn &&{ }\hskip -30pt \half (-g_{\m\a}\hat R_{\n\b}+g_{\m\b}\hat
R_{\n\a}+g_{\n\a}\hat R_{\m\b}-g_{\n\b}\hat R_{\m\a})+\fract
 16 (g_{\m\a}g_{\n\b}-g_{\n\a}g_{\m\b})\hat R\ , \eel{Weyldef}
which has the remaining 10 independent components.

The transformation rules under coordinate transformations \eqn{infcoordtrf} are as usual; all these curvature fields transform as
tensors. To see how they transform under \eqn{localscale}, first note how the connection fields transform:
 \be \hat \G_{\a\m\n}\ra(1+\l)\hat \G_{\a\m\n}+\half(\hat g_{\a\n}\pa_\m\l +\hat g_{\a\m}\pa_\n\l-\hat g_{\m\n}\pa_\a\l) + \OO(\l^2)\ , \eel{connectionscaletrf}
from which we derive
 \be \hat R_{\a\b\m\n}\ra (1+\l)\hat R_{\a\b\m\n}+\half(\hat g_{\a\n}\hat D_\b\pa_\m\l-\hat g_{\a\m}\hat D_\b\pa_\n\l-\hat
 g_{\b\n}\hat D_\a\pa_\m\l+\hat g_{\b\m} \hat D_\a\pa_\n\l)\ .
 \eel{Riemscaletrf}
From this we find how the Ricci tensor transforms:
 \be \hat R_{\m\n}\ra \hat R_{\m\n}-\hat D_\m\pa_\n\l-\half \hat g_{\m\n}\hat D^2\l\ ,\qquad \hat R\ra \hat R(1-\l)-3\hat D^2\l\ . \eel{Ricciscaletrf}
The Weyl tensor \eqn{Weyldef}, being the traceless part, is easily found to be invariant (apart from the canonical term):
 \be \hat W_{\a\b\m\n}\ra(1+\l)\hat W_{\a\b\m\n}\ . \eel{Weylscaletrf}
Since the inverse, \(\hat g^{\m\n}\), and the determinant, \(\hat g\), of the metric transform as
 \be \hat g^{\m\n}\ra(1-\l)\hat g^{\m\n}\ ;\qquad\hat g\ra (1+4\l)\hat g\ , \eel{detscaletrf}
we establish that exactly the Weyl tensor squared yields an action that is totally invariant under local
scale transformations in four space-time dimensions (remember that \(\hat g^{\m\n}\) is used to connect the
indices):
 \be\mathcal{L}=C\sqrt{-\hat g}\,\hat W_{\a\b\m\n}\hat W^{\a\b\m\n}=C\sqrt{-\hat g}(\hat R_{\a\b\m\n}\hat R^{\a\b\m\n}-
2\hat R_{\m\n}\hat R^{\m\n}+\fract 13 \hat R^2)\ ,
 \eel{Weylsqu}
which, due to the fact that the integral of
 \be \hat R_{\a\b\m\n}\hat R^{\a\b\m\n}-4\hat R_{\m\n}\hat R^{\m\n}+\hat R^2\eel{GB}
is a topological invariant, can be further reduced to
 \be\mathcal{L}=2C\sqrt{-\hat g}(\hat R_{\m\n}^2-\fract13 \hat R^2)\ , \eel{scaleinvaction}
to serve as our locally scale invariant Lagrangian.

The constant \(C\) may be any dimensionless parameter. Note that, according to Eq.~\eqn{Ricciscaletrf},
neither the Ricci tensor nor the Ricci scalar are invariant; therefore, they are locally unobservable at this
stage of the theory. Clearly, in view of Einstein's equation, \emph{matter}, and in particular its
stress-energy-momentum tensor, are locally unobservable in the same sense. This will have to be remedied at a
later stage, where we must work on redefining what matter is at scales much larger than the Planck scale.

Thus we have verified that, indeed, the action \eqn{divpart} is the only expression that we could have expected
there (apart from its overall constant) since we integrated out the scale component of the original metric
\(g_{\m\n}\). Demanding locality immediately leads to this expression.

In fact, gravity theories with this action as a starting point have been studied extensively \cite{PM}, and there
the suspicion was expressed that such theories might be unitary, in spite of the higher time derivatives in the
action. Model calculations show\cite{BM} that unitarity can be regained if one modifies the hermiticity condition,
which is equivalent to modifying the boundary conditions of functional amplitudes in the complex plane.
Effectively then, the fields become complex. Before following such a route further, we would have to understand
the underlying physics.

In Eq.~\eqn{divpart}, we arrived at the conformal action with an essentially infinite coefficient in front. Before
deciding what to do with this infinity, and to obtain more insight in the underlying physics, let us study the
classical equations that correspond to this action.

\def\ns{\!\!} To this end, consider an infinitesimal variation \(h_{\m\n}\) on the metric: \(\hat g_{\m\n}\ra \hat
g_{\m\n}+\d\hat g_{\m\n}\), \(\d\hat g_{\m\n}=h_{\m\n}\). The infinitesimal changes of the Ricci tensor and
scalar are
 \be \d \hat R_{\m\n}&=&\half(\hat D_\a \hat D_\m h^\a_\n+\hat D_\a \hat D_\n h^\a_\m -D^2 h_{\m\n}-\hat D_\m \pa_\n h^\a_\a)\
 ;\crl{Riccivary} \d \hat R&=&-h^{\a\b}\hat R_{\a\b}+\hat D_\a \hat D_\b h^{\a\b}-\hat D^2 h^\a_\a\ . \eel{scalarvary}
Using the Bianchi identity
 \be  D_\m  R^\m_{\ \n}=\half \pa_\n  R\ , \eel{Bianchi}
the variation of the Weyl action \eqn{Weylsqu}, \eqn{scaleinvaction} is then found to be
 \be &\d\LL= -2C\int\dd^nx \sqrt{-\hat g}\,h^{\a\b}\Box_{\,\a\b}^R\ ,\qquad\hbox{with}&\nn
 &\Box_{\,\a\b}^R=\hat D^2\hat R_{\a\b}-\fract13 \hat D_\a \hat D_\b \hat R-\fract16 g_{\a\b}\hat D^2 \hat R-2\hat
R^\m_\a \hat R_{\m\b}+2\hat R^{\m\n} \hat R_{\a\m\b\n}-\fract23 \hat R\hat R_{\a\b}\ .
&\eel{Weylvary}

The classical equations of motion for the Ricci tensor as they follow from the Weyl action are therefore:
 \be \Box_{\,\a\b}^R=0\ . \eel{Weylequ}
To see their most salient features, let us linearize in \(\hat R_{\m\n}\) and ignore connection terms. We get
  \be \hat R_{\m\n}-\fract16\hat R \d_{\m\n}\deff S_{\m\n}\ ; \qquad \pa_\m S_{\m\n}=\pa_\n S_{\a\a}\
;\quad\pa^2 S_{\m\n}-\pa_\m\pa_\n S_{\a\a}=0\ . \eel{eomS}
 Defining \(\l(x)\) by the equation
 \be \pa^2\l\deff -S_{\a\a}\ , \eel{Lambdadef} we find that the solution \(S_{\m\n}\) of Eq.~\eqn{eomS} can be
written as
 \be S_{\m\n}=-\pa_\m\pa_\n\l+A_{\m\n}\ ,\quad\hbox{with}\quad\pa^2 A_{\m\n}=0\ ,\ A_{\a\a}=0\ ,\ \pa_\m
 A_{\m\n}=0\ . \eel{Ssol}
From Eq.~\eqn{Ricciscaletrf} we notice that the free function \(\l(x)\) corresponds to the local scale degree
of freedom \eqn{localscale}, while the equation for the remainder, \(A_{\m\n}\), tells us that the Einstein
tensor, after the scale transformation \(\l(x)\), can always be made to obey the d'Alembert equation \(\pa^2
G_{\m\n}=0\), which is basically the field equation for the stress-energy-momentum tensor that corresponds to
massless particles\fn{Not quite, of course. The statement only holds when these particles form classical
superpositions of plane waves such as an arbitrary function of \(x-t\).}. Thus, it is not true that the Weyl
action gives equations that are equivalent to Einstein's equations, but rather that they lead to Einstein
equations with only massless matter as their source.

\newsecl{Non conformal matter}{nonconformal}\def\conf{\mathrm{conf}}

To generalize to the case that our matter fields are not conformal, the easiest case to consider is a scalar
field \(\f(x)\). Conformally invariant scalar fields are described by the action
 \be\LL^\f_\conf=-\half\sqrt{-g}(g^{\m\n}\pa_\m\f\,\pa_\n\f+\fract16 R\f^2)\ , \eel{confscalar}
where the second term is a well-known necessity for complete conformal invariance. Indeed, substituting the
splitting \eqn{gsplit} we find that the field \(\f(x)\) must be written as \(\w^{-1}\hat\f(x)\), and then
 \be \sqrt{-g}\, R\iss \w^2\bigg(\sqrt{-\hat g}\hat R\!&-&\!6\pa_\m(\sqrt{-\hat g}\,\hat g^{\m\n} {1\over \w}\,\pa_\n\w)\
+\ 6\sqrt{-\hat g} \,\hat g^{\m\n}\pa_\m\w\pa_\n\w\bigg)\ , \crl{Rsplit}
 \sqrt{- g}\,g^{\m\n}\pa_\m\f\,\pa_\n\f &=&\sqrt{-\hat g}\,\hat g^{\m\n}(\pa_\m\hat\f-{\pa_\m\w\over\w}\f)
 (\pa_\n\f-{\pa_\n\w\over\w}\f)\ ,\crl{phiactionsplit}
 \LL^\f_\conf &=& -\half\sqrt{-\hat g}(\hat g^{\m\n}\pa_\m\hat\f\pa_\n\hat\f+\fract16 \hat R\hat\f^2)\ .
 \eel{confhatscalar}
The extra term with the Ricci scalar is in fact the same as the insertion \eqn{Rterm} in Eq.~\eqn{omegaL}.
Inserting this as our matter Lagrangian leaves everything in the sections \ref{scalefint} and \ref{Weyl}
unaltered.

Now, however, we introduce a mass term:
 \be \LL^{\f,\mathrm{mass}}=\LL^\f_\conf-\half\sqrt{-g}\,m^2\f^2\ . \eel{massiveL}
After the split \eqn{gsplit}, this turns into
 \be \LL^{\f,\mathrm{mass}}=\LL^\f_\conf-\half\sqrt{-\hat g}\,m^2\w^2\hat\f^2\ . \eel{massomega}
Thus, an extra term proportional to \(\w^2\) arises in Eq.~\eqn{omegaL}. But, as it is merely quadratic in
\(\w\), we can still integrate this functional integral exactly.\fn{Note that a cosmological constant would
add a term \(C\L\w^4\) to the action, so that the \(\w\) integration can then no longer be done exactly.
Thus, there is good reason to omit the cosmological constant, but it would be premature to speculate that
this adds new views on the well-known cosmological constant problem.} At \(n\ra 4\), and remembering that we
had scaled out a factor \(6/\k^2\) in going from Eq.~\eqn{EHsplit} to Eq.~\eqn{omegaL}, the quantity \(M\) in
Eq.~\eqn{tHV} is now replaced by
 \be M=-\fract16 \hat R+\fract16{\k^2m^2}\hat\f^2\ , \eel{extraM}
and plugging it into the divergence equation \eqn{scalarpole} replaces Eq.~\eqn{divpart} by
 \be  \G^\div={\sqrt{-\hat g}\over 8\pi^2(4-n)}\left(\fract1{120}(\hat R_{\m\n}\hat R^{\m\n}-\fract13 \hat R^2)+
 \fract1{144}(\k^2m^2\hat\f^2)^2\right)\ ,\eel{divpartmassive}
where \(\k^2=8\pi G_N\). Indeed, the extra term is a quartic interaction term and as such again conformally
invariant. \(\k^2m^2\) is a dimensionless parameter and, usually, it is quite small.

The two terms in Eq.~\eqn{divpartmassive} have to be treated in quite a different way. As was explained in
Section \ref{Weyl}, the first term would require a non canonical counter term, which we hesitate to add just
like that, so it presents real problems that will have to be addressed.

This difficulty does not play any role for the second term. Its divergent part can be renormalized in the
usual way by adding a counter term representing a quartic self interaction of the scalar field. There will be
more subtle complications due to the fact that renormalization of these non gravitational interaction terms
in turn often (but not always) destroys scale invariance. As for the matter fields, these complications will
not be further considered here. Suffices to say that in some special cases, such as in supersymmetric
theories, the problems simplify.

It is important to conclude from this section that non-conformal matter does not affect the formal conformal
invariance of the effective action after integrating over the metascalar \(\w\) field. Also, the non
conformal parts, such as the mass term, do not have any effect on the dangerously divergent term in this
effective action.

\newsecl{The divergent effective conformal action}{confaction}
Let us finally address our real problem, the divergence of the effective action \eqn{divpart} as
\(n\ra 4\). This really spoils the beautiful program we outlined at the beginning of Section
\ref{Weyl}. One can imagine five possible resolutions of this problem.\fn{This section is the most
important revision in version \# 2 of this paper.}

\emph{A. Cancelation against divergences due to matter}. Besides scalar matter fields, one may have Dirac
spinors and/or gauge fields that also propagate in the conformal metric \(\hat g_{\m\n}(\vec x,\,t)\). These
also lead too divergences. Ignoring interactions between these matter fields, one indeed finds that these
fields contribute to the divergence in the effective action \eqn{divpart} as well. In fact, all these
divergences take the same form of the Weyl action \eqn{scaleinvaction}, and they each just add to the overall
coefficient. So, with a bit of luck, one might hope that all these coefficients added up might give zero.
That would certainly solve our problem. It would be unlikely that also the finite parts of the effective
action would completely cancel out, so we would end up with a perfectly conformally invariant effective
theory.

A curious problem would have to be addressed, which is that the effective action scales as the fourth power
of the momenta of the conformal \(\hat g_{\m\n}\) fields, so that there should be considerable concern that
unitarity is lost. One might hope that unitarity can be saved by observing that the theory is still based on
a perfectly canonical theory where we started off with the action \eqn{EH}.

Unfortunately, this approach is ruled out for a very simple reason: the matter fields can never
cancel out the divergence because they all contribute with the same sign! This is a rather
elaborate calculation, of a kind already carried out in the early 1970s \cite{PvN1}\cite{PvN2}. As
we are only interested in the part due to the action of scalar, spinor and vector fields on a
conformal background metric \(\hat g_{\m\n}\), we repeated the calculation and summarize its result
in the Appendix. It is found that, if the matter fields consist of \(N_0\) elementary scalar
fields, \(N_{1/2}\) elementary Majorana spinor fields (or \(\half N_{1/2}\) complex Dirac fields)
and \(N_1\) real Maxwell or Yang-Mills fields (their mutual interctions are ignored), then the
total coefficient \(C\) in front of the divergent effective action
 \be S^\eff=C\int\dd^nx{\sqrt{-\hat g}\over 8\pi^2(4-n)}\bigg(\hat R^{\m\n}\hat R_{\m\n}-\fract13 \hat R^2\bigg)\ ,
 \eel{totaldiv} is
 \be C=\fract 1{120}(1+N_0)+\fract1{40}N_{1/2}+\fract 1{10}N_1 \ . \eel{totalcoeff}
Here, the first 1 is the effect of the metascalar component \(\w\) of gravity itself. All contributions
clearly add up with the same sign. This, in fact, could have been expected from simple unitarity arguments,
but as such arguments famously failed when the one-loop beta functions for different particle types were
considered, it is preferred to do the calculation explicitly. In any case, option \emph{A} is excluded.
Although the coefficients are known from the literature \cite{PvN1}, we reproduce the details of the
calculation in the Appendix.

\emph{B. Make the integral finite with a local counter term, of the same form as Eq.~\eqn{totaldiv}, but with opposite
sign.} This is the option most physicists who are experienced in renormalization would certainly consider as the most
reasonable one. However, a combination of two observations casts serious doubts on the viability of this option. First,
in conventional theories where renormalization is carried out, this is happening in the context of a perturbation
expansion. The expression that has to be subtracted has a coefficient in front that behaves as \be {g^\a\over(4-n)^\b}\
, \eel{countercoeff} where \(g\) is a coupling strength, and the power \(\a\) is usually greater than the power \(\b\).
If we agree to stick to the limit where \emph{first} \(g\) is sent to zero and \emph{then} \(n\) is sent to 4, the
total coefficient will still be infinitesimal, and as such not cause any violation of unitarity, even if it does not
have the canonical form. This is exactly the reason why the consideration of non-canonical renormalization terms is
considered acceptable when perturbative gravity is considered, as long as the external momenta of in- and outgoing
particles are kept much smaller than the Planck value.

Here, however, Newton's constant has been eliminated, so there is no coupling constant that makes
our counter term small, and of course we consider all values of the momenta. The Weyl action is
quadratic in the Riemann curvature \(R^\a_{\,\b\m\n}\) and therefore quartic in the momenta. As
stated when we were considering option \emph{A}, one might hope that the original expression we
found can be made compatible with unitarity because it itself follows from the canonical action
\eqn{EH}. The counter term itself cannot be reconciled with unitarity. In fact, it not only
generates a propagator of the form \(1/(k^2-i\e)^2\), which at large values of \(k^2\) is very
similar to the difference of two propagators: \(1/(k^2-i\e)-1/(k^2+m^2-i\e)\), where the second one
would describe a particle with indefinite metric, but also, the combination with the total action
would leave a remainder of the form
 \be {1\over 4-n}\bigg((k^2)^{n/2} \ -\ \m^{n-4}(k^2)^2\bigg) \ \ra\ (k^2)^2\log(k^2/\m^2)\ ,
 \eel{dimcounter}
where \(\m\) is a quantity with the dimension of a mass that defines the subtraction point. The effective
propagator would take a form such as
 \be {1\over(k^2+m^2-i\e)^2\,\log(k^2/\m^2)}\ , \eel{landaughost}
which develops yet another pole, at \(k^2\approx \m^2\). This is a Landau ghost, describing something like a
tachyonic particle, violating most of the principles that one would like to obey in quantizing gravity.
 All these objections against accepting a non canonical renormalization counter term are not totally
 exclusive\cite{Mottola}, but they are sufficient reason to search for better resolutions. For sure, one would have to
 address the problems, and as yet, this seems to be beyond our capacities.

\emph{C.} An observation not yet included in an earlier version of this paper, can put our argument
in a very different light. If one follows what actually happens in conventional, perturbative
gravity, one would be very much tempted to conclude that it is incomplete: one has to include the
contribution of the \(\hat g_{\m\n}\) field itself to the infinity! Only this way, one would obtain
the complete renormalization group equations for the coefficient \(C\) in the
action~\eqn{totaldiv}. What is more, in some supergravity theories, the \emph{conformal anomaly}
then indeed cancels out to zero.\cite{Fradkin}\fn{I thank M.~Duff for this observation.} However,
arguing this way would not at all be in line with the entire approach advocated here: \emph{first}
integrate over the \(\w\) field and \emph{only then} over the fields \(\hat g_{\m\n}\). We here
discuss only the integral over \(\w\), with perhaps in addition the matter fields, and this should
provide us with the effective action for \(\hat g_{\m\n}\). If that is no longer conformally
invariant, we have a problem. Treating \(C\) as a freely adjustable, running parameter, even if it
turns out not to run anymore, would be a serious threat against unitarity, and would bring us back
to perturbative gravity as a whole, with its well-known difficulties. In addition, an important
point then comes up: how does the \emph{measure} of the \(\hat g_{\m\n}\) integral scale? This
might not be reconcilable with conformal invariance either, but also difficult if at all possible
to calculate: the measure is only well-defined if one fixes the gauge \`a la Faddeev-Popov, and
this we wish to avoid, at this stage. We neither wish to integrate over \(\hat g_{\m\n}\), nor fix
the gauge there. In conclusion therefore, we dismiss option \(C\) as well.

Therefore, yet another option may have to be considered:

\emph{D. No counter term is added at all. We accept an infinite coefficient in front of the Weyl action.} To see the
consequences of such an assumption, just consider the case that the coefficient \(K=C/(4-n)\) is simply very large. In
the standard formulation of the functional integral, this means that the quantum fluctuations of the fields are to be
given coefficients going as \(1/\sqrt{K}\). The \emph{classical} field values can take larger values, but they would
act as a background for the quantized fields, and not take part in the interactions themselves. In the limit
\(K\ra\infty\), the quantum fluctuations would vanish and only the classical parts would remain. In short, this
proposal would turn the \(\hat g_{\m\n}\) components of the metric into classical fields!

There are important problems with this proposal as well: classical fields will not react upon the
presence of the other, quantized, fields such as the matter fields. Therefore, there is no back
reaction of the metric. This proposal then should be ruled out because it violates the action =
reaction principle in physics. Furthermore, the reader may already have been wondering about
gravitons. They are mainly described by the parts of \(\hat g_{\m\n}\) that are spacelike,
traceless and orthogonal to the momentum. If we would insist that \(\hat g_{\m\n}\) is classical,
does this mean that gravitons are classical? It is possible to construct a gedanken experiment with
a device that rotates gravitons into photons; this device would contain a large stretch of very
strong, transverse magnetic fields. Turning photons into gravitons and back, it would enable us to
do quantum interference experiments with gravitons. This then would be a direct falsification of
our theory. However, the classical behavior of gravitons that we suspect, comes about because of
their interactions with the logarithmically divergent background fluctuations. If the usual
renormalization counter term of the form \eqn{totaldiv} is denied to them, this interaction will be
infinite. The magnetic fields in our graviton-photon transformer may exhibit fluctuations that are
fundamentally impossible to control; gravitons might still undergo interference, but their typical
quantum features, such as entanglement, might disappear.

Yet, there may be a different way to look at option $D$. In previous publications\cite{GtHquant},
the author has speculated about the necessity to view quantum mechanics as an \emph{emergent}
feature of Nature's dynamical laws. `Primordial quantization' is the procedure where we start with
classical mechanical equations for evolving physical variables, after which we attach basis
elements of Hilbert space to each of the possible configurations of the classical variables.
Subsequently, the evolution is re-expressed in terms of an effective Hamiltonian, and further
transformations in this Hilbert space might lead to a description of the world as we know it. This
idea is reason enough to investigate this last option further.

There is one big advantage from a technical point of view. Since \(\hat g_{\m\n}\) is now considered to be
classical, there is no unitarity problem. All other fields, both the metascalar field \(\w\) (the `dilaton')
and the matter fields are described by renormalizable Lagrangians, so that no obvious contradictions arise at
this point.

How bad is it that the action = reaction principle appears to be violated? The metric metatensor
does allow for a source in the form of an energy momentum tensor, as described in
Eqs.~\eqn{Weylequ}---\eqn{Ssol}. This, however, would be an unquantized source. We get a
contradiction if sources are described that evolve quantum mechanically: the background metric
cannot react. In practice, this would mean that we could just as well mandate that
 \be \hat g_{\m\n}=\eta_{\m\n}\ , \eel{flatspace}
in other words, we would live in a flat background where only the metascalar component of the
metric evolves quantum mechanically.\fn{This idea goes back to, among others,
Nordstr\"om\cite{Nordstrom}.}

Could a non-trivial metric tensor \(\hat g_{\m\n}\) be \emph{emergent}? This means that it is taken either to
be classical or totally flat beyond the Planck scale, but it gets renormalized by dilaton and matter fields
at much lower scales. This may be the best compromise between the various options considered. Spacetime is
demanded to be conformally flat at scales beyond the Planck scale, but virtual matter and dilaton
fluctuations generate the \(\hat g_{\m\n}\) as we experience it today. A problem with this argument,
unfortunately, is that it is difficult to imagine how dilaton fluctuations could generate a non trivial
effective metric. This is because, regardless the values chosen for \(\w(x)\), the light cones will be the
ones determined by \(\hat g_{\m\n}\) alone, so that there is no `renormalization' of the speed of light at
all. We therefore prefer the following view:

Consider a tunable choice for a renormalization counter term in the form of the Weyl action \eqn{totaldiv},
described by a subtraction point \(\m\). If \(\m\) were chosen to be at low frequencies, so at large distance
scales, then the Landau ghost, Eq.~\eqn{landaughost}, would be at low values of \(k^2\) and therefore almost
certainly ruin unitarity of the amplitudes. Only if \(\m\) would be chosen as far as possible in the
ultraviolet, this ghost would stay invisible at most physical length scales, so the further away we push the
subtraction point, the better, but perhaps the limit \(\m\ra\infty\) must be taken with more caution.

The previous version of this paper was incomplete without the following alternative option. A more mainstream
standpoint would be :

\textit{E}. The action \eqn{EH} no longer properly describes the situation at scales close to the
Planck scale. At \(|k|\approx M_{Pl}\), we no longer integrate over \(\w(k)\), which has two
consequences: a natural cut-off at the Planck scale, and a breakdown of conformal invariance.
Indeed, this would have given the badly needed scale dependence to obtain a standard interpretation
of the amplitudes computed this way. Note that, in our effective action \eqn{divpartmassive}, all
dependence on Newton's constant has been hidden in an effective quartic interaction term for the
\(\f\) field. That could have been augmented with a `natural' quartic interaction already present
in the matter Lagrangian, so we would have lost all explicit references to Newton's constant. Now,
with the explicit breakdown of conformal invariance, we get Newton's constant back.

Adopting this standpoint, it is also easy to see how a subtraction point wandering to infinity, as described in
option~$D$, could lead to a classical theory for \(\hat g_{\m\n}\). It simply corresponds to the classical limit.
Letting the subtraction point go to infinity is tantamount to forcing \(M_{Pl}\) to infinity, in which limit, of
course, gravity is classical. Only if we embrace option $D$ fully, we would insist that the physical scale is not
determined by \(M_{Pl}\) this way, but by adopting some gauge convention at a boundary at infinity. This is the
procedure demanded by black hole complementarity.

The price paid for option $E$ is, that we lost the fundamental advantages of exact conformal
invariance, which are a calculable and practically renormalizable effective interaction, and a
perfect starting point for a conformally invariant treatment of the black hole correspondence
principle as was advocated in Ref.~\cite{GtHcompl}. The idea advocated in this paper is \emph{not}
to follow option \emph{E} representing what would presumably be one of the mainstream lines of
thought. With option~\emph{E}, we would have ended up with just another parametrization of
non-renormalizable, perturbative, quantum gravity. Instead, we are searching for an extension of
the canonical action~\eqn{EH} that is such that the equivalent of the \(\w\) integration can be
carried out completely.

\newsecl{Conclusions}{concl} Our research was inspired by recent ideas about black holes \cite{GtHcompl}.
There, it was concluded that an effective theory of gravity should exist where the metascalar
component either does not exist at all or is integrated out. This would enable us to understand the
black hole complementarity principle, and indeed, turn black holes effectively indistinguishable
from ordinary matter at tiny scales. A big advantage of such constructions would be that, due to
the formal absence of black holes, we would be allowed to limit ourselves to topologically trivial,
continuous spacetimes for a meaningful and accurate, nonperturbative description of all
interactions. This is why we searched for a formalism where the metascalar \(\w\) is integrated out
first.

Let us briefly summarize here how the present formulation can be used to resolve the issue of an
apparent clash between unitarity and locality in an evaporating black hole. An observer going into
the hole does not explicitly observe the Hawking particles going out. (S)he passes the event
horizon at Schwarzschild time \(t\ra\infty\), and from his/her point of view, the black hole at
that time is still there. For the external observer, however, the black hole has disappeared at
\(t\ra\infty\). Due to the back reaction of the Hawking particles, energy (and possibly charge and
angular momentum) has been drained out of the hole. Thus, the two observers appear to disagree
about the total stress-energy-momentum tensor carried by the Hawking radiation. Now this
stress-energy-momentum tensor was constructed in such a way that it had to be covariant under
coordinate transformations, but this covariance only applies to \emph{changes} made in the
stress-energy-momentum when creation- and/or annihilation operators act on it. About these
covariant changes, the two observers do not disagree. It is the \emph{background subtraction} that
is different, because the two observers do not agree about the vacuum state. This shift in the
background's source of gravity can be neatly accommodated for by a change in the conformal factor
\(\w(x)\) in the metric seen by the two observers.

This we see particularly clearly in Rindler space. Here, we can generate a modification of the background
stress-energy-momentum by postulating an infinitesimal shift of the parameter \(\l(x)\) in
Eqs.~\eqn{Ricciscaletrf} and \eqn{Ssol}. It implies a shift in the Einstein tensor \(G_{\m\n}\) (and thus in
the tensor \(T_{\m\n}\)) of the form
 \be G_{\m\n}\ra G_{\m\n}- D_\m\pa_\n\l+g_{\m\n}D^2\l\ . \eel{Gtensorshift}
If now the transformation \(\l\) is chosen to depend only on the lightcone coordinate \(x^-\), then
 \be Q_{--}\ra G_{--}-\pa_-^2\l\ , \eel{Gminminshift}
while the other components do not shift. Thus we see how a modification only in the energy and momentum of
the vacuum in the \(x^+\) direction (obtained by integrating \(G_{--}\) over \(x^-\)) is realized by a scale
modification \(\l(x^-)\).

In a black hole, we choose to modify the pure Schwarzschild metric, as experienced by an ingoing observer, by
multiplying the entire metric with a function \(\w^2(t)\) that decreases very slowly from 1 to 0 as
Schwarzschild time \(t\) runs to infinity. This then gives the metric of a gradually shrinking black hole as
seen by the distant observer. Where \(\w\) has a non vanishing time derivative, this metric generates a non
vanishing Einstein tensor, hence a non vanishing background stress-energy-momentum. This is the
stress-energy-momentum of the Hawking particles.

Calculating this stress-energy-momentum yields an apparently disturbing surprise: it does not vanish at
apacelike infinity. The reason for this has not yet completely been worked out, but presumably lies in the
fact that the two observers not only disagree about the particles emerging from the black hole, but also
about the particles going in, and indeed an infinite cloud of thermal radiation filling the entire universe
around the black hole.

All of this is a sufficient reason to suspect that the conformal (metascalar) factor \(\w(x)\) must be
declared to be locally unobservable. It is fixed only if we know the global spacetime and after choosing our
coordinate frame, with its associated vacuum state. If we would not specify that state, we would not have a
specified \(\w\). In `ordinary' physics, quantum fields are usually described in a flat background. Then the
choice for \(\w\) is unique. Curiously, it immediately fixes for us the sizes, masses and lifetimes of all
elementary particles. This may sound mysterious, until we realize that sizes and lifetimes are measured by
using light rays, and then it is always assumed that these light rays move in a flat background. When this
background is not flat, because \(\hat g_{\m\n}\) is non-trivial, then sizes and time stretches become
ambiguous. We now believe that this ambiguity is a very deep and fundamental one in physics.

Although this could in principle lead to a beautiful theory, we do hit a real obstacle, which is, of course,
that gravity is not renormalizable. This `disease' still plagues our present approach, unless we turn to
rather drastic assumptions. The usual idea that one should just add renormalization counter terms wherever
needed, is found to be objectionable. So, we turn to ideas related to the `primitive quantization' proposal
of Ref.~\cite{GtHquant}. Indeed, this quantization procedure assumes a basically classical set of equations
of motion as a starting point, so the idea would fit beautifully.

Of course, many other questions are left unanswered. Quite conceivably, further research might turn
up more alternative options for a cure to our difficulties. One of these, of course, is superstring
theory. Superstring theory often leads one to avoid certain questions to be asked at all, but
eventually the black hole complementarity principle will have to be considered, just as the
question of the structure of Nature's degrees of freedom at distance and energy scales beyond the
Planck scale.

\section*{Acknowledgements}
The author thanks S.~Giddings, R.~Bousso, C.~Taubes , M.~Duff and P.~Mannheim for discussions, and P.~van
Nieuwenhuizen for his clarifications concerning the one-loop pole terms. He thanks R.~Jackiw for pointing out an
inaccuracy in the Introduction, which we corrected.

\appendix\newsecl{The calculation of the one-loop pole terms for scalars,\\ spinors and vectors interacting with a
background metric.}{append} The general algorithm for collecting all divergent parts of one-loop quantum
corrections in quantum field theories was formulated in Ref.~\cite{GtHalgorithm}, applied to a gravitational
background metric in Ref.~\cite{GtHVeltman}, and worked out much further in \cite{PvN2}. Here, we briefly
summarize the calculations that lead to the coefficients in Eq.~\eqn{totalcoeff}, see also Ref.~\cite{PvN1}.

Consider a quantized complex, possibly multi-component, scalar field \(\f(x)\): let its Lagrangian in a
curved background be
 \be &\LL\iss \sqrt{-g}(-g^{\m\n}D_\m\f^*\,D_\n\f)+\sqrt{g}\,\f^*(2 N^\m D_\m\f+M\f)\ ,\qquad \hbox{then}&
 \eel{backgrL} where \(g_{\m\n}\) is a 4 by 4 matrix (of course, \(g_{\m\n}\) is expected to have an inverse,
\(g^{\m\n}\)), and \(N^\m\) and \(M\) may be arbitrary, differentiable functions of the
 spacetime coordinates \(x^\m\), as well as matrices in the internal indices of the \(\f\) field. The
gradient \(D_\m\) may contain a background gauge field \(Z_\m\):
 \be D_\m\f=\pa_\m\f+Z_\m\f\ , \eel{covder}
where \(Z_\m\) may again be any function of space-time.\fn{Actually, having both a gauge field \(Z_\m\) and
an external field \(N^\m\) is redundant, but we keep them both for later convenience.}

It was derived in Ref. \cite{GtHVeltman} that the infinite component of the effective action (which we will
call the `pole term') is
 \be  \D\LL\iss{\sqrt{-g}\over\e}\Tr\left(\fract1{12}Y_{\m\n}Y^{\m\n}+\half X^2+\fract 1{60}(R_{\m\n}^2-\fract 13
 R^2)\right)\ ,  \eel{XYpole}
where we slightly modified the notation:\fn{Including an overall sign switch, since in Refs.
\cite{GtHVeltman}, \cite{PvN1} and \cite{PvN2}, the \emph{counter} term was computed.}
 \be &\e= 8\pi^2(4-n)\ ;\qquad,\qquad Z_{\m\n} \deff \pa_\m Z_\n-\pa_\n Z_\m+[Z_\m,Z_\n]\ ,& \crl{XYdefs}
 &X \iss M-N_\m N^\m-D_\m N^\m+\fract 16 R\ ,\qquad   Y_{\m\n}\iss Z_{\m\n}+D_\m N_\n-D_\n N_\m+[N_\m,N_\n] \
  ,& \nm\eel{scalarpole1} and `Tr' stands for the trace in the internal \(\f\) indices. The Lorentz
indices are assumed to me moved up and down, and summed over, using the metric \(g_{\m\n}\) in the usual way.
Naturally, the covariant derivative of the background function \(N^\m\) is defined to be
 \be D_\m N^\a=\pa_\m N^\a+\G^\a_{\m\n}Z^\n+ [Z_\m,\,N^\a]\ , \eel{covderN}

For the metascalar field \(\w\) in Section~\ref{scalefint}, we have \(M=-\fract16R\), while for the scalar
matter field in Section~\ref{nonconformal} we have \(M=m-\fract16R\), and in both these cases there is no
further gauge field or \(N^\m\) field, so \(Y_{\m\n}=0\). Since both \(\w\) and \(\f\) (Eq.~\eqn{confscalar}
were chosen to have only one single, real component, the resulting pole term has to be divided by 2. That
gives Eqs.~\eqn{scalarpole}, \eqn{divpart}, and Eq.~\eqn{divpartmassive}, leading to the first coefficient,
\(\fract1{120}\) in Eq.~\eqn{totalcoeff}.

Next, Eq.~\eqn{XYpole} can be used as a stating point to compute the pole term for Dirac and for vector
fields. First, let us consider a quantized Maxwell field \(B_\m(x)\).

We add to the Maxwell Lagrangian the gauge fixing term \(\LL_g=-\half\sqrt{-g}\,(D_\m B^\m)^2\), which, for
convenience, was chosen to be covariant for general coordinate transformations. Because of this choice, the
Faddeev Popov ghost fields \(\eta,\,\overline\eta\) now couple to the background metric. The total Lagrangian
thus becomes
 \be\LL&=&\sqrt{-g}\left(-\half(D_\m B_\n D^\m B^\n-\half B^\m(D_\n D_\m-D_\m D_\n)B^\n+\overline\eta
 D^2\eta\right) =\nn
&=&\sqrt{-g}\left(-\half(D_\m B_\n)^2-\half B_\m R^{\m\n}B_\n+\overline\eta D^2\eta\right)\ ,\eel{MaxwellL}
where indices are moved up and down using the background metric \(g_{\m\n}\), using the fact that the metric
commutes with the covariant derivative \(D_\m\).

We can now use Eq.~\eqn{XYpole} as a master equation, provided that the Lorentz indices \(\m,\n,\cdots\) of
the Maxwell field \(B\) are replaced by internal Lorentz indices \(a,b,\cdots\), using the Vierbein field
\(e_\m^a\) obeying
 \be g_{\m\n}=e^a_\m e^a_\n\ ,\qquad e_\m^a e^{a\n}=\d_\m^\n\ ,\ \hbox{etc,} \eel{vierbein}
where the summation over the internal Lorentz index \(a,b,\cdots\) is assumed to have the sign convention
\((-,+,+,+)\) in the usual way. The covariant derivative of the Maxwell field now contains the Lorentz
connection field \(A_\m^{ab}\) as a gauge field, whose curvature coincides with the Riemann tensor:
 \be D_\m B^a=\pa_\m B^a+A_\m^{ab}B^b\ ;\qquad F_{\m\n}^{ab}=\pa_\m A_\n^{ab}-\pa_\n A_\m^{ab}+
  [A_\m,A_\n]^{ab}=R^{ab}_{\ \m\n}\ . \eel{Lorentzcurvature}
Inserting the variable \(B^a\) in \eqn{backgrL}, and remembering that now it has 4 real components, we have
 \be &Z_\m^{ab}=A_\m^{ab}\ ,\qquad X_{ab} =-R_{ab}+\fract 16 R \d^{ab}\ ,\qquad Y_{\m\n}^{ab}=F_{\m\n}^{ab}\ ,& \nn
& \D\LL^B={\sqrt{-g}\over\e}\left(\fract 1{24}Y_{\m\n}^{ab}Y_{\m\n}^{ba}+\quart(R^{ab}-\fract16 R
 \d^{ab})^2+\fract 4{120}(R_{\m\n}^2-\fract13 R^2)  \right)\ =&\nn
& ={\sqrt{-g}\over\e}\left(\fract 7{60}R_{\m\n}^2-\fract 1{40}R^2\right)\ ,&\eel{vectorpole} where use was
made of the fact that the combination \eqn{GB} is a pure derivative and so can be put equal to zero.

The ghost contribution, including its sign switch, is
 \be \D\LL_g={\sqrt{-g}\over\e}\left(-\fract1{60}(R_{\m\n}^2-\fract13 R^2)-\fract1{72}R^2\right)\ , \eel{ghostpole}
and the result, when added up,
 \be \D\LL_\mathrm{Maxwell}={\sqrt{-g}\over\e}\left(\fract1{10} R_{\m\n}^2-\fract1{30} R^2 \right)\ , \eel{Maxwpole}
gives the last coefficient \(\fract1{10}\) in Eq.~\eqn{totalcoeff}.

For a derivation of the pole term coming from the Dirac fields, we can also use the master formula
\eqn{XYpole}. Here, the use of the Vierbein field will be seen to be crucial. Let \(\g^a,\ a=1,2,3,4\), be
the four Dirac \(\g\) matrices. We write
 \(\g_\m=e_\m^a\g^a\ ;\ \g_\m\g_\n=g_{\m\n}+\s_{\m\n}\ ,\ D_\m\g_\n=0\), and as the Lagrangian for a complex Dirac field
 we use
 \be \LL&=&-\sqrt{-g}\,\overline\j(\g^\m D_\m+M)\j\ , \quad D_\m\j\deff(\pa_\m+B_\m+\fract 14\s^{ab}A^{ab}_\m)\j\
  . \eel{DiracL}
The mass term and the external gauge field \(B_\m\) will actually not be used in this paper, but it is
convenient to keep them for later use, and for checking the correctness of the formalism.

Now this is a first order Lagrangian, while Eq.~\eqn{backgrL} is second order. So, instead of
Eq.~\eqn{DiracL}, we take a squared Lagrangian:
 \be &\LL=\sqrt{-g}\,\overline\j(\g D-m_1)(\g D+m_2)\j=\sqrt{-g}\left(-g^{\m\n}D_\m\overline\j D_\n\j+
 \overline\j(2N^\m D_\m +M)\j\right)\ ,& \nn
 &N_\m=\half(m_2-m_1)\g^\m\ , \qquad M=\half\s^{\m\n}G_{\m\n}+\fract18\s^{\m\n}\s^{ab}F_{\m\n}^{ab}+\g\pa
 m_2-m_1m_2\ ,& \nn
&X=m_1m_2-m_1^2-m_2^2+\half\g\pa(m_1+m_2)+\fract16 R+\fract18\s^{\m\n}\s^{ab}R_{\m\n ab}+\half\s^{\m\n}
G_{\m\n}\ ,&\nn
 &Y_{\m\n}=G_{\m\n}+\fract14\s^{ab}F^{ab}_{\m\n}+\half(\g^\m\pa_\n-\g^\n\pa_\m)(m_1-m_2)+\half(m_2-m_1)^2\s_{\m\n} \
 . & \eel{secondorderL} Here, \(G_{\m\n}\) is the covariant curl of the external \(B\) field.

Next, assuming that we have 4 complex spinor components, derive
 \be\quart\Tr(\g_\m\g_\n\g_\a\g_\b\,R_{\m\n\a\b})^2&=&4R^2\ , \nn
 \quart\Tr(\g_\a\g_\b R_{\m\n ab})^2&=&-8R_{\m\n}^2+2R^2\ , \nn
 \quart\Tr\g^\m\g^\n\g^\a\g^\b R_{\m\n\a\b}&=&-2R\ , \nn
 \quart\Tr(\s^{\m\n}G_{\m\n})^2&=&-2G_{\m\n}^2\ ,\nn
 \quart\Tr(\s_{\m\n})^2&=&-12\ .  \eel{sigmatraces} One finds
\be \quart\Tr X^2&=&m_1^4+m_2^4+3m_1^2m_2^2-2m_1m_2^3-2m_1^3m_2+\quart(\pa(m_1+m_2))^2-\half G_{\m\n}^2 \nn
&&+R^2(\fract 1{36}+\fract1{16}-\fract1{12})+(m_1m_2-m_1^2-m_2^2)(\fract13 R-\fract12R)\ , \nn
 \quart\Tr\,Y_{\m\n}Y_{\m\n}&=&G_{\m\n}^2-\half
 R_{\m\n}^2+\fract18R^2+\fract32(\pa(m_1-m_2))^2-3(m_1^2+m_2^2-2m_1m_2)^2\nn &&-\half R(m_1-m_2)^2\ . \nn
  \fract1{60}\Tr(R_{\m\n}^2-\fract13 R^2)&=&\fract 1{15} R_{\m\n}^2-\fract 1{45}R^2\ .
 \eel{DiracXY}
This, inserted into Eq.~\eqn{XYpole} for 4 complex fields, adding the Fermionic minus sign, leads to\fn{In
the derivation, it was assumed that \(m_1\) and \(m_2\) were commuting matrices, but one easily checks that
the result \eqn{doubleDiracpole} continues to hold when they do not commute.}
 \be{ }\hspace{-20pt} -\D\LL=\fract{\sqrt{-g}}{\e}\left(m_1^4+m_2^4+\pa m_1^2+\pa m_2^2-\fract23 G_{\m\n}^2+\fract16
 R(m_1^2+m_2^2)-\fract1{10}(R_{\m\n}^2-\fract13 R^2)\right)\ .  \eel{doubleDiracpole}
Notice that all cross terms containing products such as \(m_1m_2^3\) cancel out, as they must, because what
was computed here is the combined effect of two fermion species, with masses \(m_1\) and \(m_2\). One
concludes that the pole term produced by a single fermion of mass \(M\) is given by
 \be{ }\hspace{-30pt} \D\LL&=&{\sqrt{-g}\over\e}\Tr\left(-M^4-g^{\m\n}\pa_\m M\pa_\n M-\fract16 RM^2+\fract 13 G_{\m\n}^2+\fract
 1{20}(R_{\m\n}^2-\fract 13 R^2)\right)\ .  \eel{Diracpole}
A Majorana spinor counts as half a Dirac spinor, so this is how we derived the coefficient \(\fract1{40}\) in
Eq.~\eqn{totalcoeff}.

We observe that conformal invariance is obeyed throughout. If the mass terms are treated as metascalars, as
they should, we see that the Lagrangians we start off with are totally conformally invariant, and so are the
pole terms that we found. Not only does the Riemann curvature only appear in the Weyl combination,
\(R_{\m\n}^2-\fract13 R^2\), but we also see that the conformal combination \((\pa M)^2+\fract16 RM^2\)
emerges in the Dirac pole term \eqn{Diracpole}.

\end{document}